# Stackelberg Game Approach on Modeling of Supply Demand Behavior Considering BEV Uncertainty

Behrouz Azimian
*Member, IEEE*
Inamori School of Engineering
Alfred University
Alfred, NY, USA
mr.behrouz.azimian@IEEE.org

Ramin Faraji Fijani
Inamori School of Engineering
Alfred University
Alfred, NY, USA
rf8@alfred.edu

Ehsan Ghotbi
Inamori School of Engineering
Alfred University
Alfred, NY, USA
ghotbi@alfred.edu

Xingwu Wang
Inamori School of Engineering
Alfred University
Alfred, NY, USA
fwangx@Alfred.edu

*Abstract*— **The rapid progression of sophisticated advance metering infrastructure (AMI), allows us to have a better understanding and data from demand-response (DR) solutions. There are vast amounts of research on the internet of things and its application on the smart grids has been examined to find the most optimized bill for the user; however, we propose a novel approach of house loads, combined with owning a battery electric vehicle (BEV) equipped with the BEV communication controllers and vehicle-to-grid (V2G) technology. In this paper we use the Stackelberg game approach to achieve an efficient and effective optimized algorithm for the users (followers) based on time dependent pricing. We also assumed an electricity retailer company (leader) and a two-way bilateral communication procedure. The usage-based side of the game has been studied together with demand side management (DSM). Real-time pricing (RTP) from time-of-use (TOU) companies has been used for better results, and Monte Carlo simulation (MCS) handles the uncertain behavior of BEV drivers. Numerical results compared to those from the simulation show that with this method we can reshape the customer's demand for the best efficiency.**

## I. INTRODUCTION

In a traditional power grid, proper metering may decrease demand response (DR) non-scheduled loads slightly. In a smart grid, several advanced techniques can be integrated, including advanced metering infrastructure, energy management systems, distributed energy systems, intelligent electronic devices, internet of things (IoTs), and battery electric vehicles (BEVs) [1]. Smart grids are proposed as innovative power-grid systems that unite a smart metering infrastructure capable of sensing and measuring the power consumption of users [2] along with DR programs that promise solutions for improving the efficiency of future power grids [3]. Two shapes of DR have been discussed previously. First, the retailer has all the power to directly control the consumer's usage which significantly decreases the satisfaction function of the user. Second, the retailer reshapes the DR through dynamic pricing, such as critical pricing and real time pricing (RTP) [4]. The latter is our goal in this study. Due to bi-directional energy flows, price-responsive loads, intelligent electronic devices (IEDs), phasor measurement units (PMUs) etc., a quick growth of sophisticated smart metering facilities and advanced two-ways communication technologies has made managing the energy in smart grids more flexible and more stimulating [5-7]. A self-scheduling model was studied in [8] for consumers who join in day-ahead energy markets and seek to maximize their profits. Price uncertainty was studied for generation company as a leader in game theory framework [9]. It considers both demand-side reverse bids provided by an aggregator, and random outage of generating units for transmission lines [10]. Game-theoretic methods and the Stackelberg approach are widely being used to study DR and load peak shaving. For example, the authors in [11-12] proposed a Stackelberg game approach to deal with DR scheduling under load uncertainty based on real-time pricing in a residential grid. Likewise, [13-14] used a Stackelberg game approach between one power company and multiple users, challenging them to maximize their benefit with the goal of flattening the aggregated load curve. Authors in [15] attacked this problem using a bi-level hybrid multi-objective evolutionary algorithm with the purpose of optimizing the profit for the utility company. The game theoretic approach has been developed and designed for experiments (DOE), to find the clearing price in a retail electricity market with a high penetration of small and mid-size renewable suppliers [16]. However, our approach is using the mixed integer nonlinear programming to propose an accurate two-side profit achievement for the user and retailer. Having the time-of-use (TOU) pricing scheme, retailers are practically trying to charge a calculable fee for the fixed price that depends on the amount of the user's consumption during a time interval [17]. In this paper, we recommend a novel approach of flattening the peak loads by considering schedulable home appliances and BEVs, which can be charged and discharged to the grid by using grid-to-vehicle (G2V) and vehicle-to-grid (V2G) technologies. Recently, a research has been conducted on using Stackelberg game theory to coordinate electric vehicle charging schemes [18]. However, V2G technology, which makes the operator able to receive power from the BEVs, has not been practically taken into account. At last, by using an advanced two-way communication link and BEVs charging/discharging schedule, it not only guarantees the maximum profit for the utility company by peak-shaving, it also satisfies the user by elevating their utility function, which means minimizing the cost and maximizing the satisfaction. Main contributions of this paper include following key points:

- Using a price-based model to develop an efficient algorithm between retailers and users (with BEVs), and aiming for balanced supply-load by shaving peaks;
- Generating optimized results for both retailers and users with an iterative algorithm, and finding Stackelberg equilibrium (SE) to achieve optimal loads for both parties;
- Formulating 1-N leader-follower Stackelberg relationship between one retailer and N users; and adopting the RTP function of the retailer and the utility function of N users; and
- Handling uncertainty of different driving habits of BEV owners by Monte Carlo simulations (MCS).

Specifically, the rest of the paper is organized as follows. Section II discusses the system model and formulates the Stackelberg game theory. In Section III, we use data to generate load profiles of BEVs. Some key aspects of Stackelberg game theory and the logic are presented in Section IV. Results are provided in Section IV, and conclusions are drawn in Section V.

## II. SYSTEM MODEL

In Fig. 1, a model is established for one utility company and $N$ users with BEVs. These users will adjust their electricity usages by using advanced metering infrastructure and heterogeneous communication technologies, with the ultimate goals to reduce overall costs. The utility company (retailer) can provide hourly price structures to these users to encourage peak shaving and ultimately to maximize profits.

Advanced communication technologies enable the cloud computing environment and real time sensing/controlling including IoTs [19]. In this environment, an entity represents an electric load (consumer) or a prosumer such as a BEV. In the cloud environment, the real-time access to each entity is guaranteed. The system operator can categorize users into different groups. As shown in Fig.1, all prosumers can be placed in one group, and other conventional loads in other. By utilizing instantaneous two-way communication links, real-time electricity prices will be broadcasted, conventional users' consumption behaviors will be adjusted, and BEVs charging/discharging schedules can be changed. To maintain system stabilities and guarantee maximum profits, the retailer can impose load curtailment during peak hours.

### A. Retailer Company Model

In this model, the cost function for the retailer is labelled as $C_t(g_t)$, and depends on the amount of electricity provided ($g_t$) during an interval $t$; where $t \in T, T = |T|$ that is a strictly convex and monotonically increasing function.

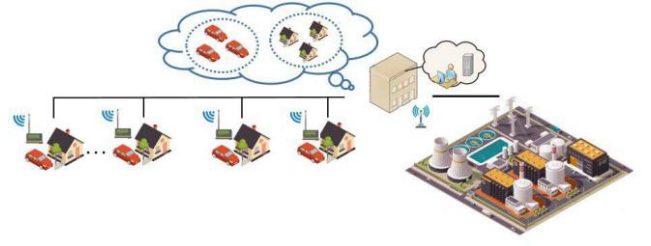

Fig. 1. System model of the retailer company and users

$$C_t(g_t) = \frac{a_t}{2} g_t^2 + b_t g_t + c_t \quad (1)$$

Where $a_t, b_t$ and $c_t$ are the generating coefficients. During different time intervals, each coefficient has a different value. Taking a differential operation against (1), one obtains the marginal cost function $C_t'$:

$$C_t'(g_t) = a_t g_t + b_t \quad (2)$$

which is the cost to produce one more unit of electricity. Such marginal cost must be lower than the actual cost in order to guarantee profits for the retailer company. So, market price equation can be defined:

$$P_t(g_t) = \lambda_t C_t'(g_t) = \lambda_t(a_t g_t + b_t), \quad \lambda_t \geq 1 \quad (3)$$

where $P_t$ is price at each time slot and $\lambda_t$ is a time-wise profit coefficient. In reference [15], a similar profit coefficient was introduced, minimized and validated. In this paper, there are 24 time intervals (slots), corresponding to 24 prices in a day. The retailer provides electricity and its price to the consumers who will decide how much to use in each time interval. Thus, a user can move his/her electricity usages to off-peak intervals to minimize the total costs. The retailer wants the most profit and the least aggregated peak load in order to avoid expensive backup generators. Flattening the demand load should be two-ways as a user has critical needs for electricity. To determine the optimal generation vector, one should minimize variations in generation, and match supply with demand. Therefore, the retailer problem can be formulated as follows:

$$\min \quad U_{RC}(g_t) = \sum_{t \in T} (g_t - \overline{g})^2 \quad (4)$$

$$\text{s.t.} \quad L_t \leq g_t \leq \min(g_t^+, L_t^{\max}) \quad (5)$$

$$L_t = \sum_{n=1}^{N} l_{n,t} \quad (6)$$

$$L_t = \sum_{i=1}^{N-1} l_{i,t} + l_{n,t} \quad i \neq n \quad (7)$$

Where $U_{RC}$ is the utility function of the retailer company and $\overline{g}$ is an average power generation during a day. $L_t$ is the summation of electricity demands of all $N$ users during the time interval t. $g_t^+$ is the maximum capacity of the retailer company's generation at the interval t, $L_t^{\max}$ is the overall upper bound of the total power demands for slot $t$. $l_{n,t}$ is the power consumed by user *n* in the interval *t* One must notice that the above-mentioned function is different from profit maximization. In (4), minimizing utility function may lead to maximized profits. In (5), the generation should always meet the total demand of all users, and be lower than the smallest threshold for generation capacity and upper load ranges. In (6) and (7), $L_t$ can be obtained by asynchronous user's adjustment of consumption, reflecting the human nature that no two users will react to real-time pricing schemes due to different needs. In (7), a user adjusts his/her usage, while others don't. In such non-simultaneous framework, no two users will negate their effects by increasing and decreasing their demands at the same time.

*B. User Model*

Each user has its own utility function, shown in (8). According to what has been explained in Section II, with the application of IoTs, the user can be an entity representing an electric load (consumer) as the first term, or a prosumer such as a BEV as the second term. Function $\psi_{n,t}(x_{n,t})$ models the satisfaction of the electricity consumers; in which $x_{n,t}$ is the general power consumption variable, $l_{n,t}$ being the residential demand and $s_{n,t}$ being BEV demand. The third term is the total amount of money to be paid by the consumers, which leads to less satisfaction as represented by a negative sign. With this, the demand side problem is formulated as follows:

$$l, s = \arg \max U_n(l_n, s_n) = \sum_{t=1}^{24} \psi_{n,t}(l_{n,t}) + \sum_{t=1}^{24} \psi_{n,t}(s_{n,t}, P_n^{BEV}) - \sum_{t=1}^{24} P_t(g_t) \cdot (l_{n,t} + s_{n,t} \cdot P_n^{BEV}) \quad (8)$$

$$\psi_{n,t}(x_{n,t}) = \omega_{n,t} x_{n,t} - \frac{\theta_n}{2} x_{n,t}^2, \quad \omega_{n,t} > 0 \quad \theta_n > 0 \quad (9)$$

s.t. $l_{n,t}^- \leq l_{n,t} \leq l_{n,t}^+$ (10)

$$\sum_{t=1}^{24} l_{n,t} = L_n \quad (11)$$

$$\sum_{t=1}^{24} s_{n,t} = T_n^{req} \quad (12)$$

$$\sum_{t=1}^{24} |s_{n,t}| \leq T_n^{max} \quad (13)$$

$$SOC_{n,T^{dep}} = BC_n \quad (14)$$

where $\omega_{n,t}$ is the preference parameter with indices *n* (user) and *t* (time), $\theta_n$ is a predetermined constant integer, $L_n$ is the total daily energy usage, $s_{n,t} \in \{-1, 0, 1\}$ is a discrete variable which is multiplied by rated battery power $P^{BEV}$ of electric vehicles and indicates BEV charging {1}, discharging {-1} status. If it is charging, it will have a positive effect on satisfaction of the utility function, and if it is discharging, it has a negative effect. Although discharging has a negative effect on satisfaction, it is losing the power obtained in the past charging periods. However, by selling discharged electricity, it has positive effect on the third term of utility function. $T_n^{req}$ is the required time to fully charge the BEV, $T_n^{max}$ is the maximum number of hours that the battery is interacting actively with the network, , $SOC_{n,T^{dep}}$ shows the BEV state of charge, and $BC_n$ is the battery capacity, $T^{dep}$ is the vehicle's departure time from the house. (9) Denotes the satisfaction function $\psi_{n,t}(x_{n,t})$ of the user *n* by consuming $l_{n,t}$ amount of electricity. (10) provides the upper and lower boundaries for electricity demand of n[th] user *n* for interval *t*. (11) expresses the temporally-coupled constraint which wouldn't reduce the total daily consumption of the user *n*, so we can apply load shedding only in terms of pick shaving and load shifting. (12) states the cumulative charging time in order to fully charge the BEV before leaving one's residence. (13) limits the maximum number of hours for a BEV connected to the grid system, either in G2V or V2G mode. Therefore, by limiting the hours that the battery is being charged or discharged, we reduce the detrimental effect of constant battery usage. In (14) the sequence of charging/discharging should be in a way that by the time that the BEV owner decides to leaves the house, the SOC should be 100%.

In the Stackelberg equilibrium context and the hierarchical process, our aim is to maximize the leader's (retailer) utility function by having the data of the follower's (user) rational reaction set (RRS). The existence of the Stackelberg equilibrium has been discussed and shown in [4]. In addition, the model in (8) is formulated as a mixed integer nonlinear programming problem, and can be solved via MATLAB-GAMS interface.

III. BEV LOAD PROFILE

To model BEV load profiles, data from National Household Travel Survey (NHTS) for rural New York State were utilized to estimate average driving mileages and optimize battery usages. The SOC of the vehicles upon their home arrival time is of crucial importance [20]. Table I shows the data according to different types of cars and different characteristics. Also, in Figs. 2-4, probability distribution functions (PDFs) for driving behavior, including arrival time, departure time, and traveled distance in rural areas in New York state has been drawn from NHTS data. The reason why we chose rural areas in New York State is that Alfred University is located in rural areas.

Moreover, configuring the time of fully charged BEV depends on the distance it travels daily, and we assume that BEV is a constant power prosumer. From Fig. 5, we obtain the required charging time based on daily driven distance [21-22].

IV. STACKELBERG GAME THEORY ALGORITHM

In Fig. 6, iterative DR algorithm is illustrated via a flow chart. At the beginning of the computation, the retailer broadcasts hourly price one hour ahead. According to (3), marginal cost function is calculated by power generation requirement and initial price broadcasted. As the relationship between the price equation and power is linear, the algorithm uses the generation value. Final prices can be found once the total power generation value is known. Afterwards, Monte Carlo simulation emulates driving behaviors of BEV owners.

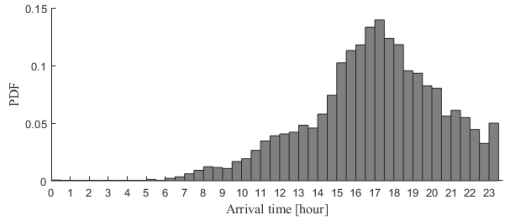

Fig. 2. Arrival time pdf for BEV owners in rural areas in New York State

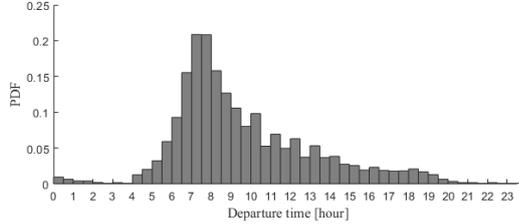

Fig. 3. Departure time pdf for BEV owners in rural areas in New York State

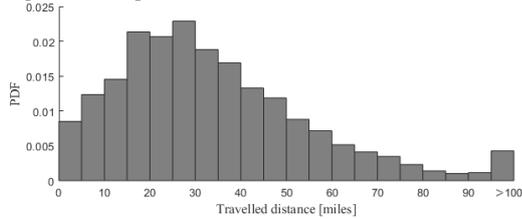

Fig. 4. Average travelled distance pdf for BEV owners in rural areas in New York State

TABLE.I   Different Characteristics of BEVs

| Vehicle Type | Commercial model | Expected Market Share [%] | Energy [kW-hr] | Rated Battery Power [kW] | Single Charge Drive-Range [mile] |
|---|---|---|---|---|---|
| Compact sedan | i3, BMW | 51.48 | 33 | 7 | 114 |
| Mid-size sedan | Model S, Tesla | 10.35 | 75 | 11.5 | 259 |
| Mid-size SUV | Model X, Tesla | 38.17 | 100 | 17.2 | 295 |

Users respond to the prices broadcasted, and shift their usages to non-peak time slots. BEV discharging features can help the retailer company to offset some loads during peak

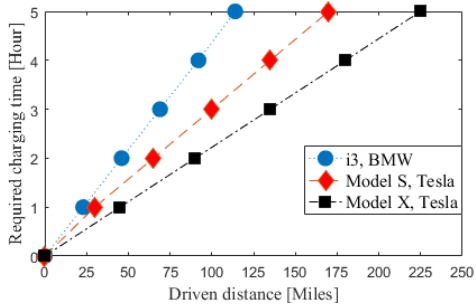

Fig. 5. Required time to fully charge a BEV with respect to distance driven and car size

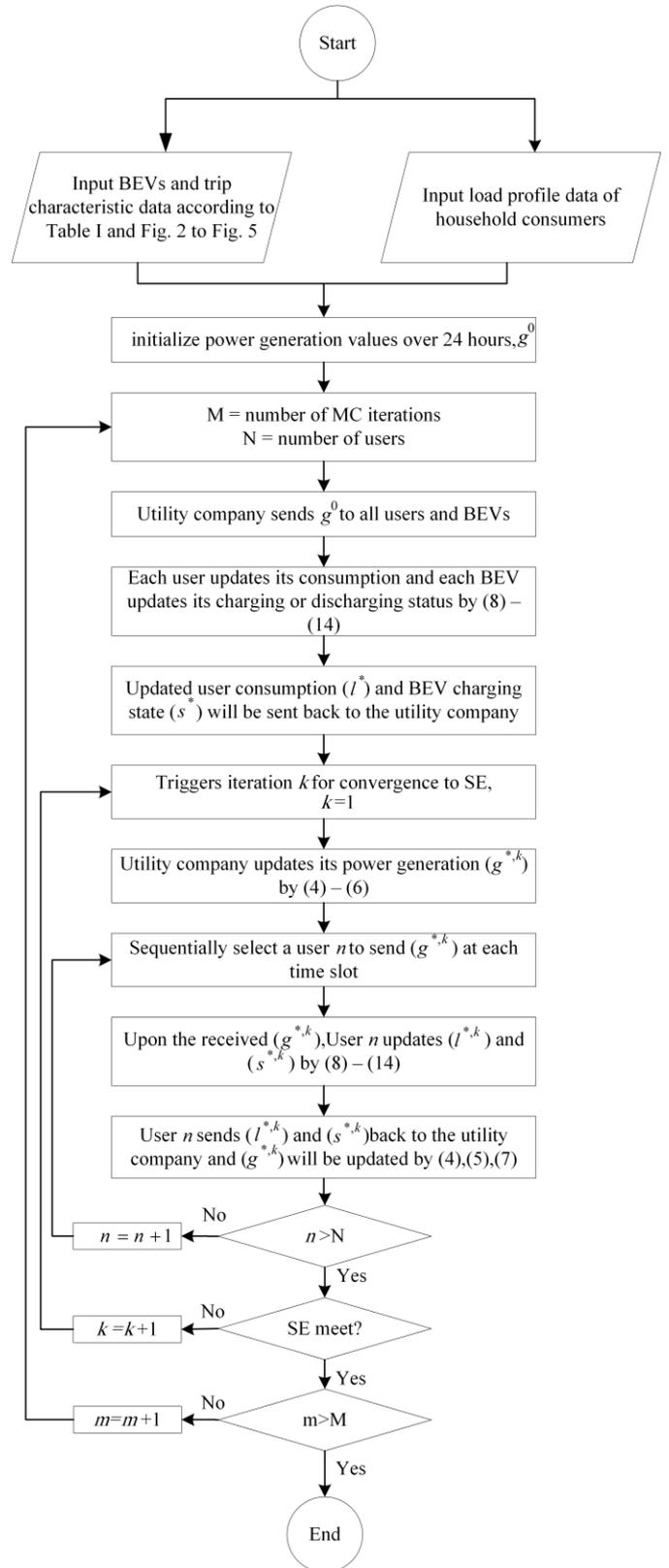

Fig. 6. Stackelberg game theory algorithm

hours. The peak-shaving and the demand aggregation can be accomplished continuously instead.

TABLE.II Satisfaction function and generation coefficients

| Group | Utility function coeff. | | Generation Coeff. | | Min Demand % | Max Demand% |
|---|---|---|---|---|---|---|
| | $\omega_{n,t}$ | $\theta_n$ | $a_t$ | $b_t$ | | |
| 1 | 5.0 | 0.1 | 0.01 (00:00-8:00) | 0.2 | 70 | 150 |
| 2 | 5.5 | 0.1 | | | 75 | 140 |
| 3 | 6.0 | 0.1 | 0.02 (8:00-24:00) | | 80 | 120 |

## V. SIMULATION AND RESULTS

Residential consumers are divided into three groups as tabulated in Table II. Group 1 corresponds to utility function coefficient $\omega_{n,t}$ of 5.0, with 50 BEVs. Groups 2 and 3 do not have BEVs, but have different utility coefficients. The minimum and the maximum values refer to the lower and upper bounds of the targeted power demands. For example, Group 1 consumption varies between 70% and 150% of its nominal load. The price coefficient is kept constant, i.e., $\lambda_t = 1.2$. In Fig. 7, load profiles are shown for Group 2 and 3; and benefits from the game theory applications can be visualized by examining a profile with or without the algorithm. In general, the load profiles are flattened due to users' participations. In comparison with Group 2, Group 3 shows less eagerness to participate as it has larger $\omega_{n,t}$. For group 1, as illustrated in Table I, managing the vehicle's V2G or G2V status is of great importance. If a large number of BEV owners begin to charge their vehicles immediately returning home, the sudden power load may be undesirable. By applying the algorithm in Fig. 6, the uncertainty related to random behaviors of BEV owners can be managed by optimizing charging and discharging activities. In Fig. 8, wide bands of load profile for BEVs are shown for each time slot in a day. Without game theory applications, the users' peak load can coincide with that of the BEV charging peaks. In Fig. 9, modified BEV load profile is obtained after applying the algorithm. The negative values represent the V2G feature of the BEVs, or discharging. After coming back to home (between 17:00 or so and midnight), most vehicles are either sending back power to the grid or in standing by mode. The charging period mainly happens between 1:00 and 6:00. Comparing the time interval between 14:00 to 16:00 in Fig. 7 and Fig. 10, more peak shaving has been achieved for Group 1 due to the presence of V2G technology. Fig. 11 shows the total generation over 24 hours for users' aggregated demand with and without implementing the DR program. The expected value at each hour is used to obtain

Table III. Numerical performance evaluation

| Scenario | Peak demand [kw] | Total Energy Usage [kwh] | Total Payments [$] | Generation Cost [$] |
|---|---|---|---|---|
| Without RTP and Without V2G | 3841 | 59092 | 36135 | 15115 |
| RTP and V2G | 2755 | 59092 | 30244 | 12661 |

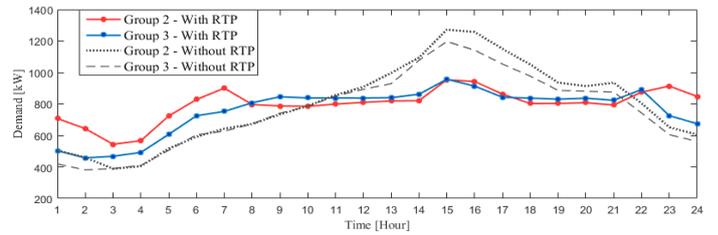

Fig. 7. Load Profiles for Groups 2 and 3 of Consumers Before and After Applying Optimization Algorithm

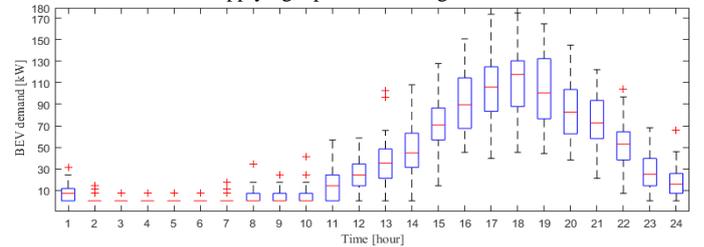

Fig. 8. Uncertain Load Profile of 50 BEVs Before Applying the Algorithm without V2G technology

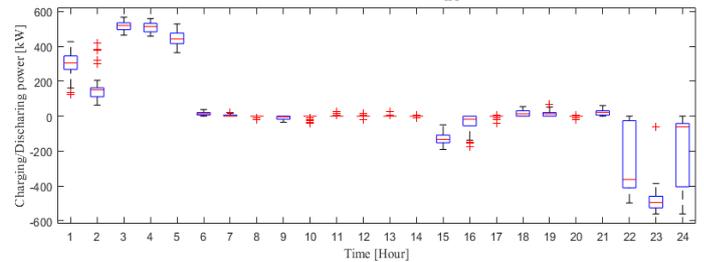

Fig. 9. Uncertain Load Profile of 50 BEVs After Applying the Algorithm with V2G technology

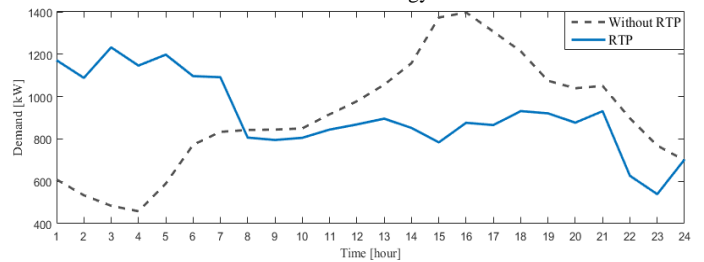

Fig. 10. Load Profiles for Group 1 of Consumers Before and After Implementing the DR program

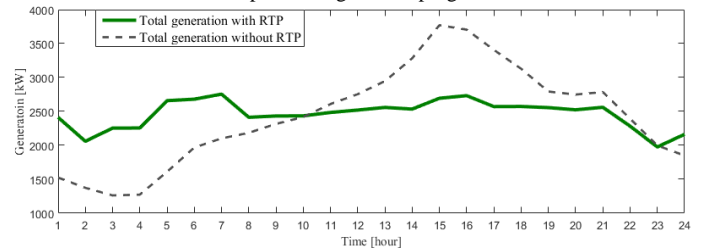

Fig. 11. Aggregated Generation Profiles Before and After Applying Optimization the Algorithm

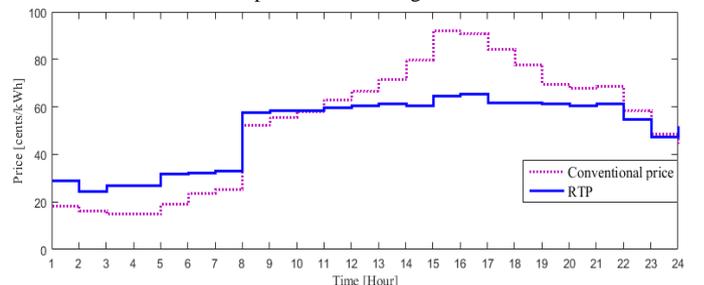

Fig. 12. Conventional and Real-Time Prices Broadcasted by Retailer

Table IV. Statistical results for generation values in [kw] considering BEV uncertainty and RTP

| Time [hour] | 1 | 2 | 3 | 4 | 5 | 15 | 16 | 17 | 18 | 19 | 20 | 21 | 22 | 23 | 24 |
|---|---|---|---|---|---|---|---|---|---|---|---|---|---|---|---|
| $g_t$ Expected value | 2412 | 2060 | 2255 | 2256 | 2659 | 2694 | 2732 | 2571 | 2573 | 2556 | 2524 | 2561 | 2283 | 1976 | 2163 |
| $g_t$ Standard deviation | 66 | 76 | 26 | 28 | 42 | 29 | 49 | 12 | 14 | 14 | 8 | 16 | 130 | 66 | 197 |

the profiles in Fig. 10 and Fig. 11. From Table IV, the standard deviation shows that more fluctuations will happen mainly from 22:00 to 24:00 and 1:00 to 5:00. Therefore, the power system operator should consider more spinning reserve after midnight; this will cause an increase in power system operation cost. Finally, the real-time prices over 24 hours that has been derived from the algorithm is compared with the conventional price in Fig. 12. The performance of the algorithm is evaluated by two scenarios (with and without RTP and V2G) from various aspects. The numerical results are shown in Table III. The peak demand is significantly decreased by almost 1 MW. Total energy usage has not been changed, which proves that the retailer cannot reduce the whole amount of energy that a user expects to consume per day. This reality is shown in (11). Using algorithm in Fig. 6, the total user payments decreased about 16%, which demonstrates the efficiency of game theory application. Additionally, the generation costs of both scenarios can be less than the total payments, i.e., making profits for the retailer.

## CONCLUSION

Using game theory algorithm, one can reduce costs to consumers and potentially reshape generating profiles. In this paper, a model included one retailer and *N* users (with some of them owning BEVs). In particular, an optimized approach can shave peaks with DR managements. The game theory and Stackelberg equilibrium have been utilized to illustrate an algorithm for three different user groups. By comparing the results with and without DR managements, the efficiency of the algorithm can be shown. Using MCS in MATLAB-GAMS, the stochastic behavior of BEVs are simulated. According to the statistical analysis, such algorithm based on the game theory can reduce the peak loads. To extend this study, having multiple retailer companies can be taken into consideration.